# Variation of the relative biological effectiveness with fractionation in proton therapy: analysis of prostate cancer response

Juan Pardo-Montero[1,2,*], Miguel Pombar[2], Antonio Gómez-Caamaño[3], Simona Giordanengo[4], Isabel González-Crespo[1,5]

1. Group of Medical Physics and Biomathematics, Instituto de Investigación Sanitaria de Santiago (IDIS), Santiago de Compostela, Spain.

2. Department of Medical Physics, Complexo Hospitalario Universitario de Santiago de Compostela, Spain.

3. Department of Radiation Oncology, Complexo Hospitalario Universitario de Santiago de Compostela, Spain.

4. Istituto Nazionale di Fisica Nucleare-Torino Section, Torino, Italy.

5. Department of Applied Mathematics, Universidade de Santiago de Compostela, Spain.

**Corresponding author:** Juan Pardo-Montero, Grupo de Física Médica e Biomatemáticas, Instituto de Investigación Sanitaria de Santiago (IDIS), Servizo de Radiofísica e Protección Radiolóxica, Hospital Clínico Universitario de Santiago, Trav. Choupana s/n, 15706, Santiago de Compostela (Spain); E-mail: juan.pardo.montero@sergas.es

**Acknowledgements:** J.P-M. acknowledges the hospitality of Istituto Nazionale di Fisica Nucleare-Torino section, where part of this work was carried out, and the funding provided by Instituto de Salud Carlos III (ISCIII) for such research stay (BA21/00012). This project has received funding from Ministerio de Ciencia e Innovación, Agencia Estatal de Investigación and FEDER, UE (grant number PID2021-128984OB-I00). This project has received funding from Xunta de Galicia, Axencia Galega de Innovación (grant number IN607D 2022/02).

**Abstract**

Background: In treatment planning for proton therapy a constant Relative Biological Effectiveness (RBE) of 1.1 is used, disregarding variations with linear energy transfer, clinical endpoint or fractionation.

Purpose: To present a methodology to analyze the variation of RBE with fractionation from clinical data of tumor control probability (TCP) and to apply it to study the response of prostate cancer to proton therapy.

Methods and Materials: We analyzed the dependence of the RBE on the dose per fraction by using the LQ model and the Poisson TCP formalism. Clinical tumor control probabilities for prostate cancer (low and intermediate risk) treated with photon and proton therapy for conventional fractionation (2 Gy(RBE)×37 fractions), moderate hypofractionation (3 Gy(RBE)×20 fractions) and hypofractionation (7.25 Gy(RBE)×5 fractions) were obtained from the literature and analyzed aiming at obtaining the RBE and its dependence on the dose per fraction.

Results: The theoretical analysis of the dependence of the RBE on the dose per fraction showed three distinct regions with RBE monotonically decreasing, increasing or staying constant with the dose per fraction, depending on the change of $(\alpha, \beta)$ values between photon and proton irradiation (the equilibrium point being at $(\alpha_P/\beta_P)=(\alpha_X/\beta_X)(\alpha_X/\alpha_P)$). An analysis of the clinical data showed RBE values that decline with increasing dose per fraction: for low risk RBE≈1.124, 1.119, and 1.102 for 1.82 Gy, 2.73 Gy and 6.59 Gy per fraction (physical proton doses), respectively; for intermediate risk RBE≈1.119, and 1.102 for 1.82 Gy, and 6.59 Gy per fraction (physical proton doses), respectively. These values are nonetheless very close to the nominal 1.1 value.

Conclusions: In this study, we have presented a methodology to analyze the RBE for different fractionations, and we used it to study clinical data for prostate cancer and evaluate the RBE versus dose per fraction. The analysis shows a monotonically decreasing RBE with increasing dose per fraction, which is expected from the LQ formalism and the changes in $(\alpha, \beta)$ values between photon and proton irradiation. However, the calculations in this study have to be considered with care as they may be biased by limitations in the modeling assumptions and/or by the clinical data set used for the analysis.

# Introduction

Proton therapy is undergoing rapid expansion **[1]**. Proton beams allow a better dose conformation to the tumor than photon beams, and the higher ionization density (higher linear energy transfer, LET) of proton beams leads to increased cell killing. This increased cell killing of particle beams (protons, carbon ions, ...) is accounted for through the Relative Biological Effectiveness (RBE). The RBE is the ratio of photon to proton/ion dose producing the same biological effect. In treatment planning for proton therapy a constant RBE of 1.1 is used, disregarding variations with LET, clinical endpoint or fractionation **[2]**.

There is experimental evidence showing that the RBE increases with LET, a dependence that has also been extensively studied in radiobiological modeling **[2-8]**. This leads to spatial variations of the RBE within the tumor and normal tissues, with maximum RBE values at the distal edge of the spread-out Bragg peak (SOBP) and the fragmentation tail after the Bragg peak **[2, 9]**. This may have important consequences on the toxicity in organs lying in that region. The RBE is also expected to depend on the fractionation and the α/β of the tumor and normal tissue. These dependencies and their potential clinical implications have been analyzed with radiobiological modeling but are less well studied experimentally **[10-13, 2]**.

Several studies analyzed clinical data aiming at detecting the effects of a variable RBE, mostly on toxicity **[9, 12-14]**. In this work, we have investigated the effect of a variable RBE with fractionation via tumor control probability. We built on modeling studies that thoroughly studied this dependence to develop a methodology for analyzing clinical data to obtain the dependence of the RBE on the fractionation. We applied this methodology to clinical data for prostate cancer, a type of cancer treated with photon and proton treatments with standard fractionation as well as moderately and strongly hypofractionated treatments **[15]**.

# Methods and materials

### Radiobiological modeling and relative biological effectiveness

We relied on the LQ model to fit the dose-response **[16]**. The surviving fraction of tumor cells after a treatment delivering $n$ fractions of dose $d$ ($D=nd$) is:

$$-\log SF = E = \alpha D + \beta dD - \lambda \max(0, T - T_k) \quad (1)$$

where $\alpha$ and $\beta$ are the linear and quadratic LQ parameters, and the last term models the proliferation as an exponential function with rate $\lambda$ after a kick-off time $T_k$.

We modeled tumor control probability (TCP) using the *Poisson LQ* formalism [17],

$$TCP = \exp(-N_0 SF) \quad (2)$$

where $N_0$ is the number of clonogenic cells in the tumor.

If we introduce the biologically effective dose [18], BED,

$$BED = D + \frac{dD}{(\alpha/\beta)} - \frac{\lambda}{\alpha}\max(0, T - T_k) \quad (3)$$

we can write equation (2) as:

$$TCP = \exp(-N_0 \exp(-\alpha BED)) \quad (4)$$

The RBE is defined as the ratio of conventional and proton doses leading to an isoeffect:

$$RBE = \left(\frac{D_X}{D_p}\right)_{isoeffect} \quad (5)$$

Tumor response to a proton treatment is characterized by the parameters ($\alpha_p$, $\beta_p$), while the response to a conventional photon treatment is characterized by the parameters ($\alpha_X$, $\beta_X$). We considered the same proliferation parameters for proton therapy and conventional radiotherapy.

If we model the effect with equation (1), the RBE can be obtained by solving the following equation

$$E_p = \alpha_p n d_p + \beta_p n d_p^2 - \lambda \max(0, T - T_k) = \alpha_X n d_p RBE + \beta_X n (RBE d_p)^2 - \lambda \max(0, T - T_k), \quad (6)$$

to obtain the dependence of the RBE on the fractional dose as [10]:

$$RBE = \frac{(\alpha_X/\beta_X)}{2 d_p}\left(-1 + \sqrt{1 + \frac{4}{(\alpha_X/\beta_X)\alpha_X}(d_p \alpha_p + d_p^2 \beta_p)}\right) \quad (7)$$

The variation of the RBE with fractional dose, $d_p$, depends on the changes in radiosensitivity between photon ($\alpha_X$, $\beta_X$) and proton radiation ($\alpha_p$, $\beta_p$). From equation (7) one can prove (shown in the Appendix) that the RBE would be a constant function with the dose per fraction, $RBE(d)=RBE_0$, if:

$$\alpha_p = RBE_0 \alpha_X; \quad \beta_p = RBE_0^2 \beta_X \tag{8}$$

More generally, it can be shown that the RBE increases/decreases with the dose per fraction according to (see the Appendix):

$$(\alpha_p/\beta_p) > (\alpha_X/\beta_X)(\alpha_X/\alpha_p) \Rightarrow \text{RBE monotonically decreases with dose per fraction}$$

$$(\alpha_p/\beta_p) = (\alpha_X/\beta_X)(\alpha_X/\alpha_p) \Rightarrow \text{RBE does not depend on dose per fraction} \tag{9}$$

$$(\alpha_p/\beta_p) < (\alpha_X/\beta_X)(\alpha_X/\alpha_p) \Rightarrow \text{RBE monotonically increases with dose per fraction}$$

## Methodology to analyze clinical RBE values

The methodology to analyze clinical data and obtain the RBE and its dependence on fractionation is as follows:

**Step 1:** Fit the model (equations (1) and (2)) to clinical dose-response data for different fractionations for photon treatments to obtain best-fitting values for ($\alpha_X$, $\beta_X$, $\lambda$, $T_k$).

**Step 2:** Use the parameters obtained in Step 1 to obtain the RBE for each fractionation. From equations (1) and (2) we can write:

$$SF_p = \frac{-\log(TCP_p)}{N_0} = \exp\left(-\alpha_X D_p RBE - \beta_X d_p D_p RBE^2 + \lambda \max(0, T-T_k)\right) \tag{10}$$

which leads to the following quadratic equation:

$$(\beta_X d_p D_p) RBE^2 + (\alpha_X D_p) RBE - \lambda \max(0, T-T_k) + \log(-\log(TCP_p)/N_0) = 0 \tag{11}$$

This equation can be easily solved to obtain the RBE as:

$$RBE=\frac{-\alpha_X D_p+\sqrt{(\alpha_X D_p)^2+4\beta_X d_p D_p\left(\lambda\max(0,T-T_k)-\log(-\log(TCP_p)/N_0)\right)}}{2\beta_X d_p D_p} \quad (12)$$

The number of clonogenic cells in the tumor, $N_0$, in equation (2) was not considered as a fitting parameter and was instead set to $N_0=4.5\times10^5$ for LR and $N_0=3\times10^6$ for IR prostate cancer, according to the analysis of Pedicini *et al.* **[19]**.

**Step 3:** Fit the *(RBE, $d_p$)* data obtained in Step 2 by equation (7) to obtain best-fitting *($\alpha_p$, $\beta_p$)* values.

**Step 4:** Use the values obtained in Steps 3 and 1 to reconstruct the curve *(RBE, $d_p$)*.

## Clinical data

We analyzed the response of prostate cancer because this cancer is treated with photon and proton therapy, with different fractionations with doses per fraction ranging from < 2 Gy to 10 Gy, and there are many published studies reporting tumor control probability that can be used for the purpose of this work **[15]**. For conventional radiotherapy (photons), we relied on the data set presented in **[20]**, while for proton therapy we performed a search in Pubmed to find studies reporting control. Tumor control probability was defined as *freedom from biochemical relapse* at 5 years post-treatment, both for photons and protons.

When comparing both data sets, we found equivalent schedules for conventional fractionation (2 Gy × 37 fractions), moderate hypofractionation (3 Gy × 20 fractions) and hypofractionation (7.25 Gy × 5 fractions) **[21-32]**. For proton therapy, these doses are *effective doses*, which we will denote by Gy(RBE), including an RBE of 1.1. Information on the schedules used in this work is reported in Table 1. Data reported in Table 1 correspond to low risk (LR) and intermediate risk (IR) prostate cancer, which were analyzed separately. We also found equivalent schedules for high risk prostate cancer, but only for 2 Gy/fraction, and therefore they were discarded from the analysis.

**Table 1:** Clinical data used in this study. For each study, we report dose per fraction (d), number of fractions (n), overall treatment time (OTT), number of patients (N) and tumor control probability (TCP). For each fractionation, a single control value was calculated weighted by the number of patients. For intermediate risk patients, androgen deprivation therapy (ADT) is usually employed as part of the treatment. The percentage of patients receiving ADT is presented within parentheses, even if it was not used in this study. The reference of each study is included.

| Low risk patients | | | | |
|---|---|---|---|---|
| **Fractionation** | | | **conventional** | **proton** |
| **n** | **d (Gy(RBE))** | **OTT (d)** | | |
| 37 | 2 | 50 | N=157, TCP=0.968 **[21]** | N=335, TCP=0.987 **[22]**<br>N=215, TCP=0.970 **[23]**<br>N=7, TCP=1 **[24]**<br>**weighted: N=557, TCP=0.981** |
| 20 | 3 | 27 | N=164, TCP=0.968 **[21]** | N=146, TCP=0.979 **[25]** |
| 5 | 7.25 | 9 | N=189, TCP=0.990 **[26]**<br>N=45, TCP=0.977 **[27]**<br>N=67, TCP=0.940 **[28]**<br>N=324, TCP=0.965 **[29]**<br>N=61, TCP=0.944 **[30]**<br>N=365, TCP=0.965 **[31]**<br>**weighted: N=1051, TCP=0.967** | N=121, TCP=0.969 **[32]** |
| **Intermediate risk patients** | | | | |
| **Fractionation** | | | **conventional** | **proton** |
| **n** | **d (Gy(RBE))** | **OTT (d)** | | |
| 37 | 2 | 50 | N=779, TCP=0.867 (100%) **[21]** | N=894, TCP=0.910 (41%) **[22]**<br>N=520, TCP=0.910 (53%) **[23]**<br>**weighted: N=1414, TCP=0.910** |
| 5 | 7.25 | 9 | N=215, TCP=0.875 (11%) **[26]**<br>N=153, TCP=0.913 (22%) **[29]**<br>N=50, TCP=0.942 (38%) **[30]**<br>N=346, TCP=0.952 (0%) **[31]**<br>**weighted: N=764, TCP=0.922** | N=158, TCP=0.900 (17%) **[32]** |

## Statistical methods

The fitting of the TCP model to tumor control probability data (*step 1*) was performed by using the maximum likelihood methodology, assuming binomial statistics for the reported control values. The fitting of RBE versus dose data (*step 3*) was performed through the minimization of the squared differences. For the fitting of IR data, because only two fractionations were available, proliferation was ignored ($\lambda=0$).

Confidence intervals (CI) of the computed parameters were obtained by using a Monte Carlo approach **[33]**. The procedure was as follows:

i) random tumor control probabilities for each fractionation (conventional and proton therapy) were sampled from a binomial distribution *B(N, p, k)* (with *N* the number of patients, *p* the clinical TCP, and *k* the number of successes). Uncertainties in the number of clonogenic cells were also considered by sampling the number of clonogens from a gamma distribution qualitatively matching the 95% confidence intervals reported in Pedicini *et al.* **[19]** ([$2.5\times10^5$, $11\times10^5$] and [$0.8\times10^6$, $8.5\times10^6$] for LR and IR, respectively);

ii) those randomly generated values were used in the methodology presented in the previous section (Steps 1 to 3) to obtain best-fitting values for $\alpha_X$, $\beta_X$, $\lambda$, $T_k$, $\alpha_p$, $\beta_p$ and RBEs;

iii) steps i-ii were repeated 10000 times to accumulate statistics;

iv) 95% CIs for each parameter were obtained by discarding extremal values;

v) p-values were computed from confidence intervals following the methodology presented by Altman & Bland in **[34]**.

These methods were implemented and run in Matlab (Mathworks, Natick, MA).

# Results

In Table 2 we show best-fitting parameter values obtained with the methodology presented in this study. The fit of dose-response data for conventional radiotherapy leads to values of $\alpha_X \approx 0.15$ Gy$^{-1}$ (0.13 Gy$^{-1}$), $\beta_X \approx 0.04$ Gy$^{-2}$ (0.05 Gy$^{-2}$), and $\alpha_X/\beta_X \approx 3.5$ Gy (2.7 Gy) for LR (IR) prostate cancer. These fits are shown in Figure 1.

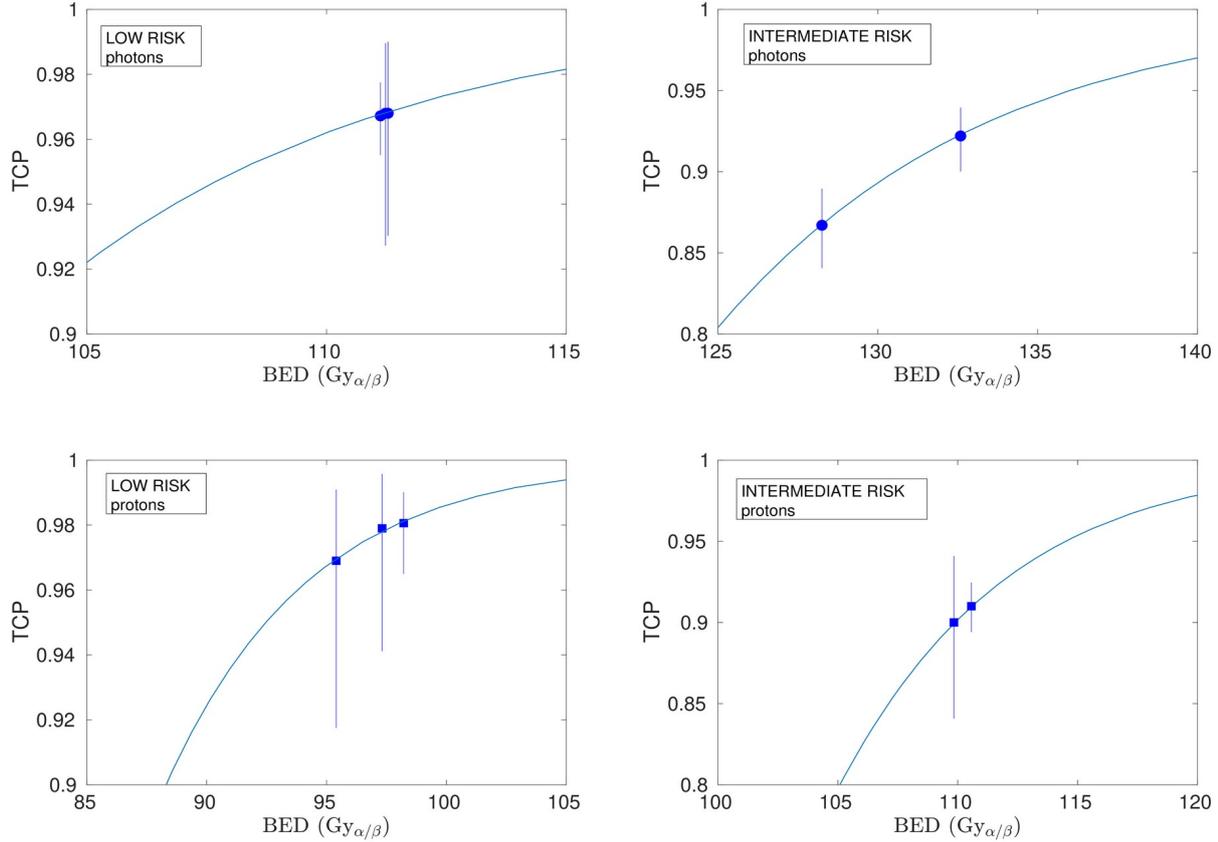

**Figure 1:** Tumor control probability (TCP) versus Biologically Effective Dose (BED) for photon and proton therapy of low and intermediate risk tumors. Solid squares and circles represent the clinical data, with error bars showing the 95% confidence intervals calculated by using the binomial distribution. The curves show the modeled dose-response. For photon treatments, the curves were obtained by fitting the clinical data to equations 1-2, as detailed in *Step 1* of the methodology. For protons, the curves were obtained by using the ($\alpha_p$, $\beta_p$) values that best fit the RBE dependence on the fractionation (*Step 3* of the methodology).

When using the best-fitting parameters shown above to obtain the RBE values for each fractionation for LR and IR (*Step 2* of the methodology) we obtained values very close to the nominal 1.1 value, which nonetheless showed a modest decline with increasing dose per fraction: RBEs for LR are 1.124, 1.119, and 1.102 for 1.82 Gy, 2.73 Gy and 6.59 Gy per fraction (physical proton doses), respectively; RBEs for IR are 1.119, and 1.102 for 1.82 Gy, and 6.59 Gy per fraction, respectively.

RBE values versus dose per fraction and the fit to the functional dependence obtained from the LQ model (equation 7) are shown in Figure 2. This fit led to values of $\alpha_p \approx 0.17$ Gy$^{-1}$ (0.16 Gy$^{-1}$), $\beta_p \approx 0.05$ Gy$^{-2}$ (0.06 Gy$^{-2}$), and $\alpha_p/\beta_p \approx 3.5$ Gy (2.8 Gy) for low risk (intermediate risk) prostate cancer.

In Table 2 we also present 95% confidence intervals for each parameter, obtained from the Monte Carlo method used to quantify uncertainties. Because of the limited number of schedules and the limited number of patients per schedule, confidence intervals are quite wide, especially for low risk data. For low risk prostate cancer, the observed decrease of RBE with increasing dose was not significant ($p$=0.71), but for intermediate risk cancer the RBE obtained at 7.25 Gy(RBE) was significantly lower than that obtained at 2 Gy(RBE) ($p$=0.02). Changes in $\alpha/\beta$ between conventional and proton irradiations were not significant. This is illustrated in Figure 3.

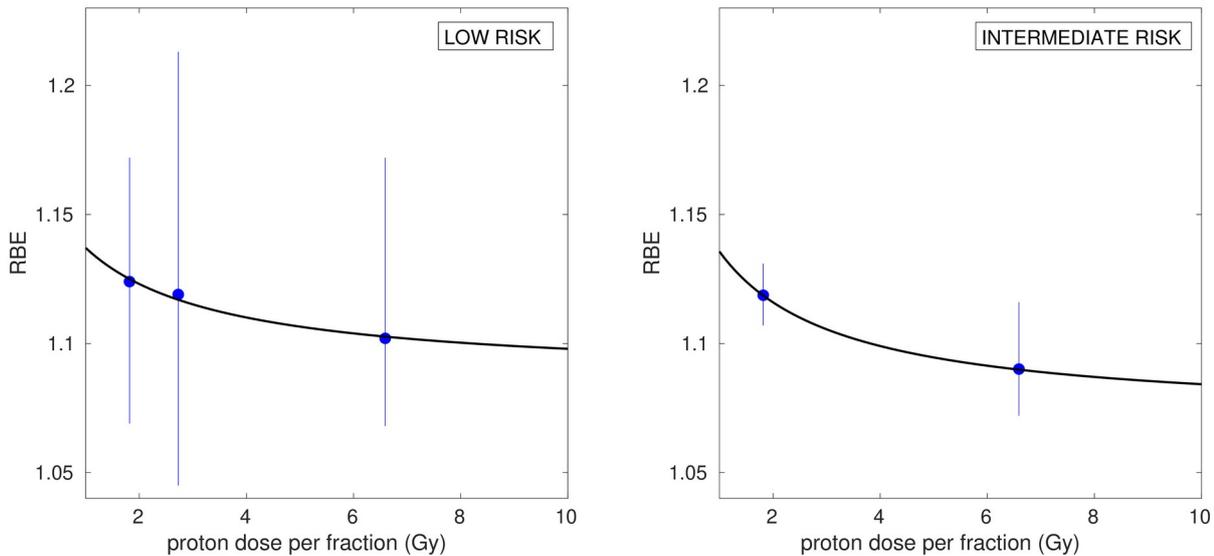

**Figure 2:** Calculated RBE versus the proton dose per fraction for low and intermediate risk prostate cancer. Error bars represent 95% confidence intervals. The solid lines represent the best fits to the RBE dependence on the fractionation expected from the LQ model (equation 7).

**Table 2:** Best-fitting parameter values and 95% confidence intervals.

| Parameter | LOW RISK | | INTERMEDIATE RISK | |
|---|---|---|---|---|
| | **Best-fitting value** | **95% CI** | **Best-fitting value** | **95% CI** |
| $\alpha_X$ [Gy$^{-1}$] | 0.148 | (0.129, 0.192) | 0.132 | (0.122, 0.141) |
| $\beta_X$ [Gy$^{-2}$] | 0.042 | (0.036, 0.046) | 0.048 | (0.045, 0.052) |
| $\lambda$ [day$^{-1}$] | 0.226 | (0.008, 0.250) | - | - |
| $T_k$ [day] | 46.77 | (30.33, 50.00) | - | - |
| $(\alpha/\beta)_X$ [Gy] | 3.513 | (2.896, 5.263) | 2.728 | (2.561, 2.901) |
| $\alpha_p$ [Gy$^{-1}$] | 0.173 | (0.143, 0.216) | 0.156 | (0.145, 0.168) |
| $\beta_p$ [Gy$^{-2}$] | 0.050 | (0.041, 0.062) | 0.055 | (0.051, 0.061) |
| $(\alpha/\beta)_p$ [Gy] | 3.478 | (2.388, 5.512) | 2.825 | (2.512, 3.083) |
| RBE (2 Gy(RBE)) | 1.124 | (1.069, 1.172) | 1.119 | (1.107, 1.131) |
| RBE (3 Gy(RBE)) | 1.119 | (1.045, 1.213) | - | - |
| RBE (7.25 Gy(RBE)) | 1.102 | (1.068, 1.172) | 1.090 | (1.072, 1.116) |

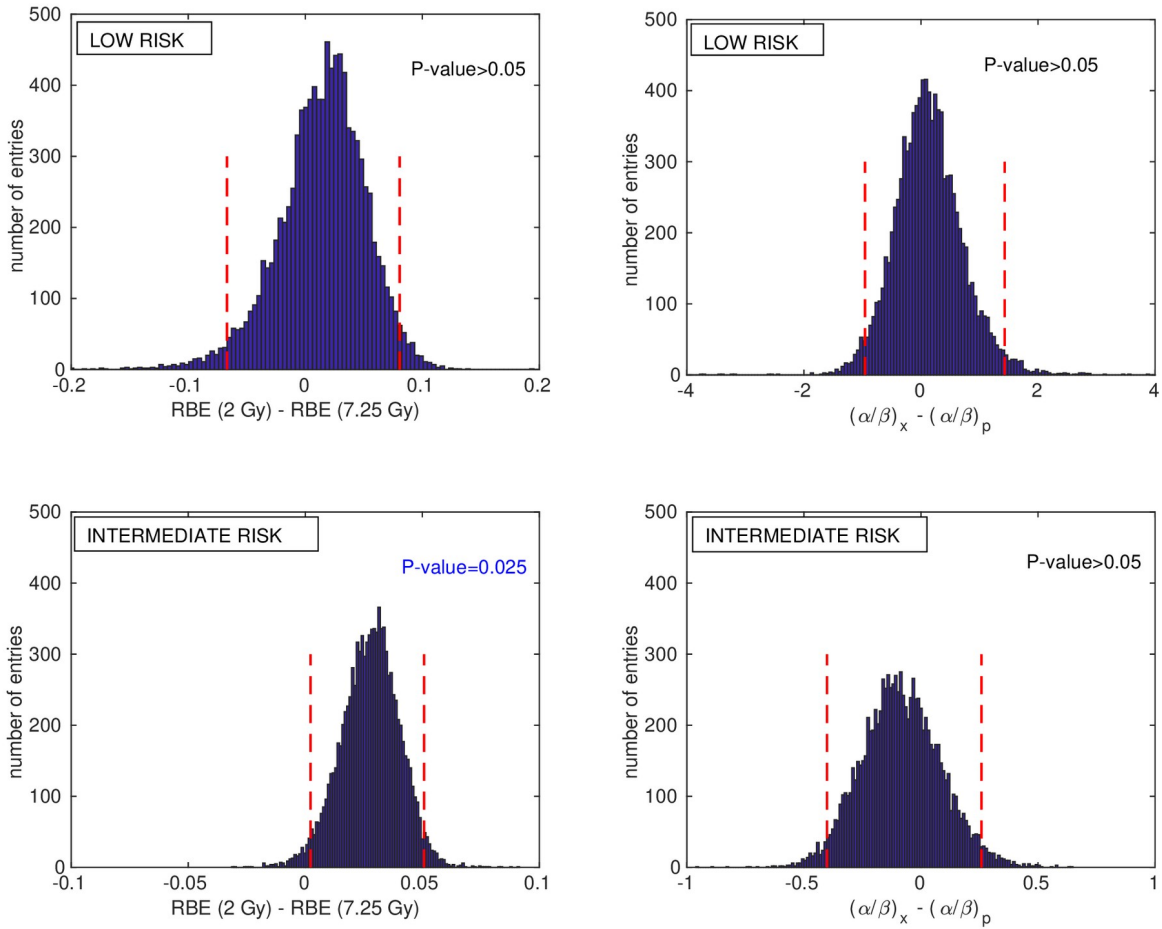

**Figure 3:** Distribution of differences between RBE at 2 Gy(RBE)/fraction and 7.25 Gy(RBE)/fraction and between $(\alpha_X/\beta_X)$ and $(\alpha_p/\beta_p)$ obtained from the Monte Carlo evaluation of uncertainties for low risk and intermediate risk prostate cancer. The dashed lines represent the 95% confidence intervals.

Classical radiobiological modeling based on the LQ model predicts a variable RBE with fractionation that depends on the $\alpha/\beta$ characterizing the response of the tumor to both radiation qualities (equation 9). The RBE may decrease or increase with increasing dose per fraction, with a boundary at $(\alpha_P/\beta_P)=(\alpha_X/\beta_X)(\alpha_X/\alpha_P)$. This behaviour is shown in Figure 4.

Theoretical and experimental work suggests that the higher radiobiological effect of protons results in $\alpha_P > \alpha_X$ and $\beta_P \approx \beta_X$ **[10].** Therefore, it is expected that $(\alpha_P/\beta_P) > (\alpha_X/\beta_X)$, and according to equation (9) the RBE of protons is expected to decrease with increasing fraction dose, as observed in our analysis.

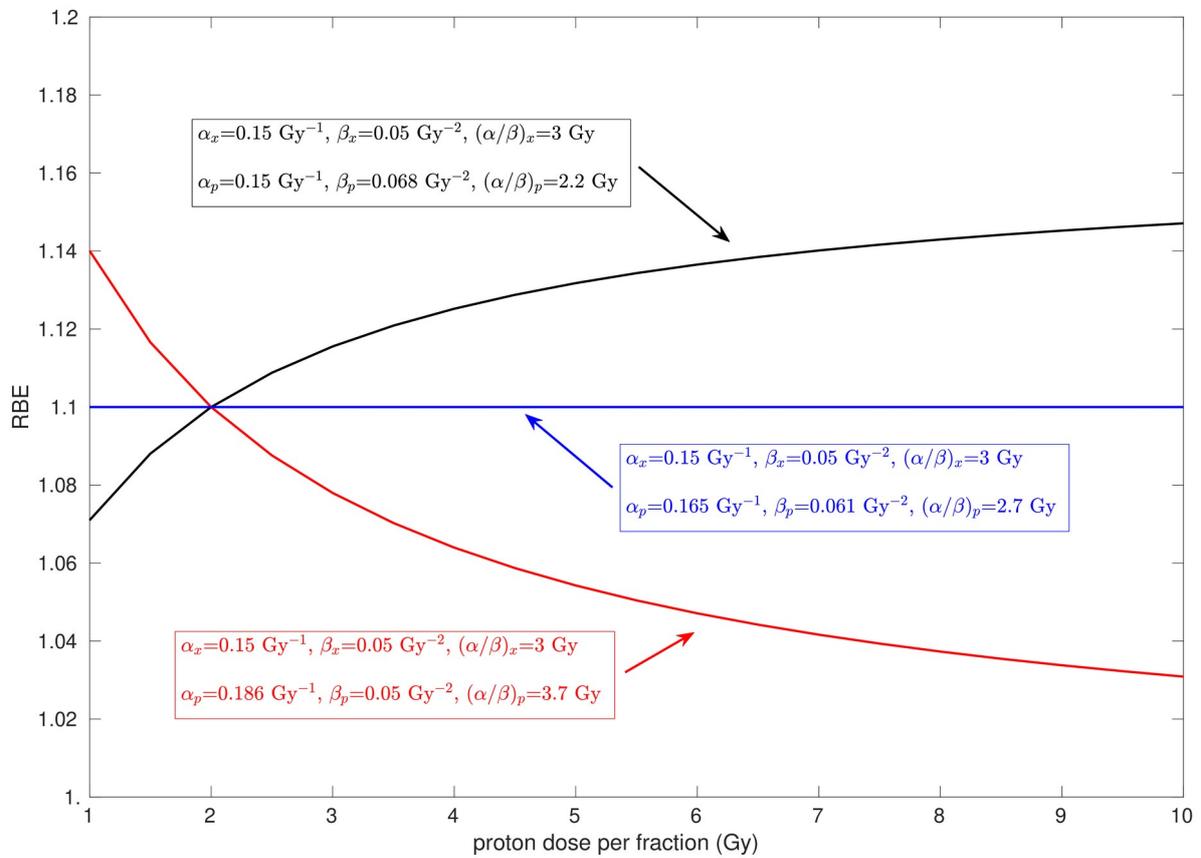

**Figure 4:** Theoretical calculation of the RBE versus proton dose per fraction based on the LQ formalism. Three regimes are observed: i) $(\alpha_P/\beta_P) > (\alpha_X/\beta_X)(\alpha_X/\alpha_P)$ monotonically decreasing RBE; ii) $(\alpha_P/\beta_P)=(\alpha_X/\beta_X)(\alpha_X/\alpha_P)$ constant RBE; iii) $(\alpha_P/\beta_P) < (\alpha_X/\beta_X)(\alpha_X/\alpha_P)$ monotonically increasing RBE. The $\alpha$ and $\beta$ values were selected to illustrate the three regimes of a variable RBE with fractionation while having RBE =1.1 at 2 Gy for all cases.

# Discussion

The RBE is expected to depend on the LET of the radiation and on the clinical endpoint. In particular, there is evidence of increasing RBE with increasing LET of protons. This leads to a higher RBE at the distal edge of the SOBP in single field treatments, which may especially affect the toxicity in organs lying in that region. In multi-field IMPT, the variation of RBE with LET can create a heterogeneous RBE distribution within the tumor and organs at risk **[2]**. The LET dependence of the RBE was not considered in this work, but has been thoroughly investigated by other studies **[9-14]**.

The RBE can also depend on the dose per fraction and the $α/β$ of the tumor, although this has been less well studied experimentally **[2, 10]**. Classical radiobiological modeling based on the LQ model predicts a variable RBE with fractionation that depends on the $α/β$ characterizing the response of the tumor to both radiation qualities.

The dependence on the dose per fraction was studied in this work. In particular, we presented a methodology to analyze the RBE based on tumor control probability data for different fractionations. This methodology was then used to analyze clinical data for prostate cancer and to obtain the RBE for different dose levels per fraction. We relied on the TCP formulation based on the *Poisson LQ* formalism, but ignoring the averaging of (heterogeneous) radiosensitivities. This was done to avoid a complex expression for the TCP that would not allow a closed-form expression for the RBE. It is well known that if averaging of radiosensitivities is not included, the *Poisson LQ* formalism produces TCP curves that are far too steep around $D_{50}$ (TCP=0.5) **[17]**. However, this is not expected to be a limitation for the particular case of prostate cancer investigated in this study, because the evaluated control rates are very high (>90%) and lie in a very narrow interval.

Our analysis only considered the dependence of the RBE on fractionation, and ignored the LET dependence. Since a spread-out Bragg peak exhibits a variable LET, the RBE values determined in our study may also be affected, and a proper analysis would have to consider the combined dependence of the RBE on the LET and dose. This however was beyond the scope of our work. In addition, if tumor control probability is mostly dominated by low dose areas as generally considered, one might argue that the effect of the distal high LET on the TCP would be modest.

The analysis of clinical data of dose-response for prostate showed a monotonically decreasing RBE with increasing dose per fraction, which was expected from the modeling of the RBE with the LQ formalism. The observed decrease of RBE with increasing dose was not significant for low risk prostate cancer. For intermediate risk cancer, the RBE obtained at 7.25 Gy(RBE) was found to be significantly lower than that obtained at 2 Gy(RBE) ($p$=0.025). However, even the latter result must be considered with care, as it may be biased by limitations in the modeling assumptions as well as by limitations in the clinical data set used for the analysis. The latter includes the limited number of studies as well as the assumption of similarly adequate margins or dose constraints. Finally, it has to be emphasized that the main purpose of presenting this analysis was to illustrate the application of the methodology rather than to drawing definitive conclusions on the proton RBE.

**Appendix**

Let us write equation (7) as:

$$RBE(d_p) = \frac{a}{d_p}\left(-1 + \sqrt{1 + bd_p + cd_p^2}\right) \tag{A1}$$

where $a$, $b$, and $c$ are constants given by:

$$a = \frac{(\alpha_X/\beta_X)}{2}; \quad b = \frac{4\alpha_p}{(\alpha_X/\beta_X)\alpha_X}; \quad c = \frac{4\beta_p}{(\alpha_X/\beta_X)\alpha_X} = (\beta_p/\alpha_p)b \tag{A2}$$

We can write equation A1 as:

$$RBE(d_p) = \frac{-a}{d_p} + \frac{a}{d_p}\sqrt{1 + bd_p + cd_p^2} \tag{A3}$$

The condition under which the RBE becomes independent of the fractional dose, $d_p$, is then described by:

$$\frac{a}{d_p}\sqrt{1 + bd_p + cd_p^2} = \frac{a}{d_p} + RBE_0 \tag{A4}$$

Taking squares on both sides we get

$$\frac{a^2 b}{d_p} + a^2 c = \frac{2 a RBE_0}{d_p} + RBE_0^2 \quad (A5)$$

and comparing the coefficients in $d_p$ leads to

$$ab = 2 RBE_0; \quad a\sqrt{c} = RBE_0 \quad (A6)$$

Taking the values of a, b and c (A2) we obtain the condition for a dose-independent RBE reads:

$$\alpha_p = RBE_0 \alpha_X; \quad \beta_p = RBE_0^2 \beta_X \quad (A7)$$

To investigate the dose dependence of the RBE for the case when this condition is not fulfilled, we consider the derivative of the RBE function (A1):

$$\frac{d\,RBE}{d(d_p)} = RBE'(d_p) = \frac{a\left(2\sqrt{1+bd_p+cd_p^2} - bd_p - 2\right)}{2 d_p^2 \sqrt{1+bd_p+cd_p^2}} \quad (A8)$$

Because *a, b, c* and $d_p$ are ≥0, we will have $RBE'(d_p)<0$ (and $RBE(d_p)$ monotonically decreasing) if:

$$2\sqrt{1+bd_p+cd_p^2} - bd_p - 2 < 0 \quad (A9)$$

which leads to:

$$\begin{aligned}
&2 + bd_p > 2\sqrt{1+bd_p+cd_p^2} \\
\Rightarrow &4 + b^2 d_p^2 + 4 bd_p > 4\left(1+bd_p+cd_p^2\right) \\
\Rightarrow &\frac{b^2}{4} > c
\end{aligned} \quad (A10)$$

Taking the values of *b* and *c* from (A2) we obtain:

$$(\alpha_p/\beta_p) > (\alpha_X/\beta_X)(\alpha_X/\alpha_p) \quad (A11)$$

We can generalize this result to obtain:

$$\begin{aligned}
&(\alpha_p/\beta_p) > (\alpha_X/\beta_X)(\alpha_X/\alpha_p) \Rightarrow \text{RBE monotonically decreases with dose per fraction} \\
&(\alpha_p/\beta_p) = (\alpha_X/\beta_X)(\alpha_X/\alpha_p) \Rightarrow \text{RBE does not depend on dose per fraction} \\
&(\alpha_p/\beta_p) < (\alpha_X/\beta_X)(\alpha_X/\alpha_p) \Rightarrow \text{RBE monotonically increases with dose per fraction}
\end{aligned} \quad (A12)$$

intermediate-risk prostate cancer – 5 year outcomes. Int J Radiation Oncol Biol Phys 2021;110:1090-1097

**[33]** Joint Committee for Guides in Metrology. JCGM 101: Evaluation of Measurement Data - Supplement 1 to the "Guide to the Expression of Uncertainty in Measurement" - Propagation of Distributions Using a Monte Carlo Method, 2008

**[34]** Altman DG, Bland JM. How to obtain the P value from a confidence interval. BMJ 2011;343:d2304